\documentclass{IOS-Book-Article}

\usepackage{mathptmx}
\usepackage{soul}\setuldepth{article}
\usepackage{amsfonts}
\usepackage{graphicx} 
\usepackage{algorithm}
\usepackage{tabto}
\usepackage{algpseudocode}
\usepackage{float}
\usepackage{booktabs}
\usepackage{dblfloatfix}
\usepackage{amsmath}
\usepackage{algpseudocode}
\usepackage[dvipsnames]{xcolor}

\usepackage[numbers]{natbib}
\def\hb{\hbox to 11.5 cm{}}

\begin{document}

\pagestyle{headings}
\def\thepage{}
\begin{frontmatter}              

\title{Improving AI-generated music with user-guided training}

\markboth{}{April 2025\hb}

\author[A, B]{\fnms{Vishwa Mohan} \snm{Singh}%
\thanks{Corresponding author: Vishwa Mohan Singh, Vishwa.Singh@campus.lmu.de}},
\author[A, B]{\fnms{Sai Anirudh} \snm{Aryasomayajula}},
\author[A, B]{\fnms{Ahan} \snm{Chatterjee}},
\author[A, B]{\fnms{Beste} \snm{Aydemir}}
and
\author[A]{\fnms{Rifat Mehreen} \snm{Amin}}

\runningauthor{V.M. Singh et al.}
\address[A]{Institut für Statistik, Ludwig-Maximilians-Universität München}
\address[B]{Institut für Informatik, Ludwig-Maximilians-Universität München}

\begin{abstract}
AI music generation has advanced rapidly, with models like diffusion and autoregressive algorithms enabling high-fidelity outputs. These tools can alter styles, mix instruments, or isolate them. Since sound can be visualized as spectrograms, image-generation algorithms can be applied to generate novel music. However, these algorithms are typically trained on fixed datasets, which makes it challenging for them to interpret and respond to user input accurately. This is especially problematic because music is highly subjective and requires a level of personalization that image generation does not provide.
In this work, we propose a human-computation approach to gradually improve the performance of these algorithms based on user interactions. The human-computation element involves aggregating and selecting user ratings to use as the loss function for fine-tuning the model. We employ a genetic algorithm that incorporates user feedback to enhance the baseline performance of a model initially trained on a fixed dataset. The effectiveness of this approach is measured by the average increase in user ratings with each iteration. In the pilot test, the first iteration showed an average rating increase of 0.2 compared to the baseline. The second iteration further improved upon this, achieving an additional increase of 0.39 over the first iteration.
\end{abstract}

\end{frontmatter}
\markboth{April 2025\hb}{April 2025\hb}
\pagestyle{empty}

\section{Introduction}
Advancements in deep learning have revolutionized generative AI, enabling significant progress in music generation. From Variational Autoencoders (VAEs) and Generative Adversarial Networks (GANs)\cite{NIPS2014_5ca3e9b1}, these architectures have been used in a range of applications, including generating images, music, and human-like text. Their effectiveness has been propelled significantly with transformers and other attention-based architectures \cite{vaswani2017attention}. Early models like OpenAI’s GPT and DALL-E \cite{ramesh2021zero}, as well as Google's Parti \cite{yu2022scaling}, leveraged autoregressive models based on VQ-VAEs \cite{van2017neural}. However, recently, latent diffusion models \cite{rombach2022high} have become much more prevalent due to their speed and computational efficiency, which has been further improved by works like Diffusion Transformers \cite{peebles2023scalable} to improve the fidelity. In music generation, approaches have evolved from traditional methods such as Markov chains \cite{Hill2011MarkovMG} and rule-based systems to neural networks like Recurrent Neural Networks (RNNs)\cite{Sherstinsky_2020}, LSTMs \cite{6795963}, Autoencoder-based models like MusicVAE \cite{roberts2019hierarchical} and OpenAI's Jukebox \cite{dhariwal2020jukebox}, and GAN based methods like GANSynth\cite{engel2019gansynth}. One of the most recent developments in this has been with methods using latent diffusion models, such as Riffusion \cite{Forsgren_Martiros_2022}, Noise2Music \cite{huang2023noise2music}, and Moûsai \cite{schneider2023mousai}. These are emerging as fast and powerful tools in this field, which uses spectrograms of audio as the image target. Despite these advancements, current music generation models face two key limitations: 
\begin{itemize}
    \item Firstly, they often rely on limited text input, making it difficult to capture complex musical requirements.
    \item Secondly, they are typically trained on static datasets, lacking the ability to incorporate user-specific preferences, which are critical given music’s inherently subjective nature.
\end{itemize}
To address these challenges, we propose a solution based on a combination of iterations of stable diffusion and human computation. Instead of text, users provide songs to guide the model’s generation. Furthermore, the model leverages human computation by gathering user feedback to continuously refine the model with a weighted loss through online learning. This approach enhances personalization and aims to bridge the gap between static data-driven models and dynamic, user-driven creativity.

\begin{figure*}[h]
  \centering
    \includegraphics[scale=0.66]{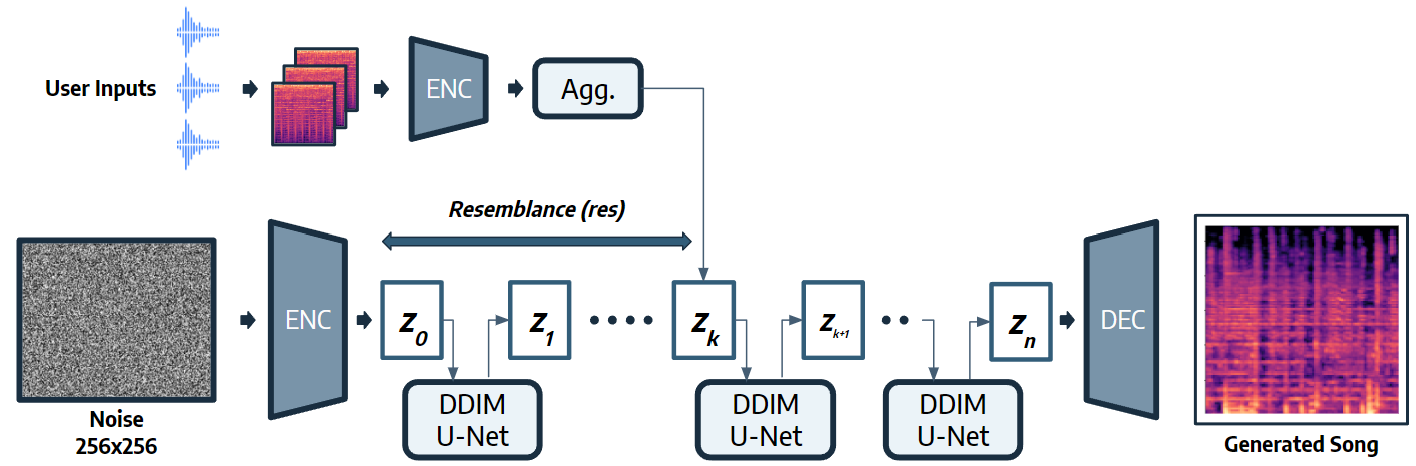}
\caption{Architecture of the stable diffusion model to generate songs that resemble the user inputs.}
    \label{fig:arch}
\end{figure*}

\section{Human Computation System}
This method qualifies as a human computation system as described in \cite{10.1145/1978942.1979148}, where humans and computers collaborate to solve problems that neither could solve independently. The primary motivation for this HC system is to enhance generation quality with human input, similar to other hybrid intelligence systems \cite{dell2024toward, mao2024hi}.. In this section, we describe how parts of our system fit under the human computation framework.

\subsection{HC System Classification} \label{hccclass}
The following classification dimensions from \cite{10.1145/1978942.1979148} are employed to analyze our system: \textit{motivation, quality control, aggregation, human skill, process order, task-request cardinality.}

\begin{itemize}
    \item \textbf{Motivation:} The primary motivator for this HC system is entertainment, similar to the ESP Game. Users are also motivated by the opportunity to contribute to the development of a popular generative model. Other motivators, such as pay, altruism, and implicit work, are not used.
    
    \item \textbf{Quality Control:} Quality control is ensured by giving the same generated song to multiple users, creating an output agreement. Adversarial behavior is also addressed during this process. In addition to explicit user ratings, implicit feedback, such as listening time, serves as a sanity check. For instance, if a user gives maximum ratings to all songs but spends minimal listening time, their inputs are adjusted accordingly. Additional quality control methods, such as statistical filtering, multilevel review, and reputation systems from \cite{10.1145/1978942.1979148}, may be applied based on system performance.
    
    \item \textbf{Aggregation:} User feedback is aggregated to improve the system. Newly generated songs are added to the training set of the music generation model, following the embedding and aggregation methods outlined previously. "Successful" samples are used iteratively to guide the model's learning.
    
    \item \textbf{Human Skill:} The system does not rely on specific human skills but rather on human taste in music, meaning all users contribute equally. However, measures are taken if adversarial behavior is detected.
    
    \item \textbf{Process Order:} According to \cite{10.1145/1978942.1979148}, an HC system involves three roles: the requester, worker, and computer. In this case, users rate the generated songs as workers, while requesters are the end-users who benefit from the music generation system. The process order is defined as Computer $\rightarrow$ Worker $\rightarrow$ Computer. It should be noted that requesters and workers are the same users, playing different roles within the process order.
    
    \item \textbf{Task-Request Cardinality:} The task-request cardinality is many-to-many, as multiple users will provide ratings to train the system.
\end{itemize}

\subsection{HC System Success Criteria} \label{successcrit}
The performance of the HC system can be evaluated through metrics that assess user engagement and the extent to which user inputs contribute to model improvement.

\begin{itemize}
    \item \textbf{Average Rating Increase}: With each iteration, the average rating of songs from users should increase, provided the iteration includes sufficient training samples to drive improvement.

    \item \textbf{Sufficient Training Pairs}: As more songs are generated, the accumulation of training samples increases, allowing the model to improve in subsequent iterations.
\end{itemize}

\begin{figure*}[h]
    \centering
    \includegraphics[scale=0.72]{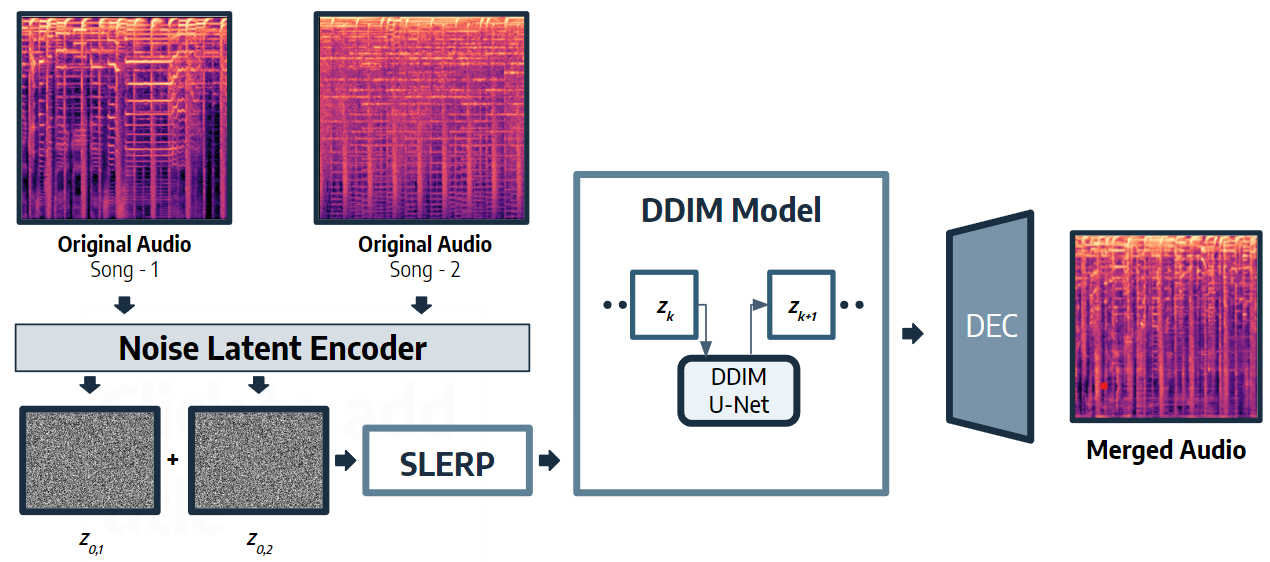}
    \caption{Flow of aggregation method for two songs. This could be extended to more songs by performing SLERP iteratively.}
    \label{fig:merge}
\end{figure*}

\section{Human Computation Algorithm and Rating Collection}
\subsection{Algorithm}\label{sec:algorithm}
 The goal of the algorithm is to identify the most suitable audio for training our model. Our objective is to include high-quality audio in the target dataset based on user listening times and ratings, ultimately improving the appeal of the generated audio.

The algorithm is divided into three key components:

\begin{itemize}
    \item Collecting Song Inputs: User-provided songs are converted to their frequency domain representation. These are encoded and aggregated to produce a unified representation of all inputs.

    \item Collecting Ratings:  When a user requests new music, there is a 25\% chance they receive a previously generated song from the dataset ($z_0$) and a 75\% chance they will receive a newly generated song ($\hat{z}$) based on combined input data. User ratings and listening times are then recorded for further tasks.

    \item Updating Targets using Ratings: If the newly generated song ($\hat{z}$) receives a better rating than the existing song ($z_0$), it is added to the training dataset to further improve the model.
\end{itemize}

The core of the algorithm is to identify newly generated songs that users prefer over the original training targets. These higher-rated songs are then incorporated into the training set, enhancing the model's ability to produce more appealing music in future iterations. As we gather more user data and refine the training process, the model will continuously improve, generating music that sounds progressively better to users.

Algorithm~\ref{alg_1} shows the algorithm for the aforementioned steps. Note that $\Gamma()$ represents the similarity function for the encodings, $G$ represents the music-generating algorithm (written as a distribution), and $R$ represents the rating function:
\begin{algorithm*}
\caption{Music Roll Out and Rating Collection Algorithm}\label{alg_1}
\begin{algorithmic}[1]
\Require $S_1, S_2, S_3$ \Comment{Upto 3 songs taken from the user}

\State $z_1,z_2,z_3 \gets Encoder(S_1,S_2,S_3)$
\State $z_{agg} \gets aggregate(z_1,z_2,z_3)$
\State $z_0  \gets \underset{z_0'}{\arg\max}\; \Gamma(z_{agg}, aggregate(z_1', z_2', z_3')) \;\;\;\;\forall z_1', z_2', z_3' \leftrightarrow z_0' \in DS_{train}$  \Comment{Most similar generation in training data}
\\


\State $ratings \gets Dict()$
\State $rating_{z_0} \gets None$
\While{User prompts to generate music}
\State $\varepsilon \gets U(0,1)$
\If{$\varepsilon < 0.25$ and $rating_{z_0}$ is $None$}
\State $\hat{z} \gets z_0$ \Comment{Giving the original target}
\State $rating_{z_0} \gets R(\hat{z})$
\Else
\State sample $\hat{z} \sim G(z;z_{agg})$ \Comment{Giving a generated song}
\State append $\hat{z} : R(\hat{z})$ to $ratings$
\EndIf
\EndWhile\\

\For{$\hat{z},rating_{\hat{z}}$ in $ratings$} \Comment{Updating new training targets}
\If{$rating_{\hat{z}} > rating_{z_0}$ }
\State Update$(z_1,z_2,z_3 \leftrightarrow \hat{z})$ 
\EndIf
\EndFor

\end{algorithmic}
\end{algorithm*}

\subsection{The Rating (R) and Update function}
The rating function depends on multiple implicit factors like the relative listening time, the coherence and similarity of the chosen songs, and explicit user ratings. The explicit user ratings are collected from the user interface using a star or score-based system and are normalized before factoring them into the rating function. Similarly, the listening time is also collected and normalized. The overall rating is defined as the following function:
\begin{equation}
\begin{split}
    R(\hat{z}) = F(\Gamma(z_1, z_2, z_3), \;\Gamma(\hat{z},aggregate(z_1, z_2, z_3)),\\ \;User\_Rating(\hat{z}),\;Listen\_Time(\hat{z}))
\end{split}
\end{equation}
where $\Gamma()$ represents the similarity function for the encodings.\\
The current rating function takes an unweighted average of all the factors mentioned above. To prevent redundant samples from filling up $DS_{train}$, we set up a similarity threshold for all the songs that would be stored and removed from the violators in a purge cycle. In this cycle, older and obsolete samples are also removed.

\section{Song Merging and Generation}
\subsection{Music Generation with Stable Diffusion} We generate new music using Denoising Diffusion Implicit Models (DDIM) \cite{song2020denoising}, which improves upon the speed of Denoising Diffusion Probabilistic Models \cite{ho2020denoising} by using a non-markovian sampling process. Although the model is trained to go from 256x256 resolution noise to the spectrogram, the user-given conditioning songs are fed to the model as an intermediate latent in the middle of a diffusion denoising step, as shown in Figure \ref{fig:arch}. The distance from step 0 determines the resemblance of the output to the conditioning music. Unlike traditional LDM, no conditioning is fed through cross-attention to the U-Net. We control the model's learning by calculating a confidence score $\omega$, the weighted mean of the song’s rating and listen time, normalized between 0-1. The loss function used to guide the learning is a weighted version of the DDIM target as follows: \begin{equation} L(\theta) = \omega \cdot\mathbb{E}[||\epsilon - \epsilon_{\theta}(\sqrt{\Bar{\alpha_t}}z_0 - \sqrt{1-\Bar{\alpha_t}}\epsilon,t)||^2] 
\label{eq:loss}\end{equation}

\subsection{Music Aggregation Function} To merge multiple song encodings into one, we use Spherical Linear Interpolation (SLERP) \cite{jafari2014spherical} and Denoising Diffusion Implicit Models (DDIM). Our process involves: \begin{itemize} \item \textbf{Step 1:} The spectrograms of all songs are converted into the $z_0$ latent vector of the DDIM model using its scheduling parameters.
\item \textbf{Step 2:} All the latent noise vectors are merged into a single representation with SLERP. 
\item \textbf{Step 3:} The merged latent is passed through the DDIM model to get the final latent vector. \end{itemize} Using this process allows us to create a single latent that closely resembles the input audio spectrograms and is still comprehensible. Figure \ref{fig:merge} illustrates the aggregation method.

\subsection{Similarity Calculation} For the similarity function $\Gamma()$, a VQ-VAE \cite{van2017neural} suits us better than the VAE used in the DDIM model. This is because the codebook allows us to compute a single latent representation $\mathbf{e}_{k^*}$ from any number of input songs. With a codebook with $K$ embeddings, $v_j$ as the encoder output of VQ-VAE for song $j$, $k^*$ for $n$ songs can be computed as follows:
\begin{equation}
 k^* = \arg\min_{k \in \{1, \dots, K\}} \sum_{j=1}^n \|\mathbf{v}_j - \mathbf{e}_k\|_2^2
\end{equation} 

Adding a classifier head allows us to incorporate information about the genre in the latent representation using deep metric learning \cite{kaya2019deep}. The similarity between $n$ songs is calculated as follows: \begin{equation}
    \Gamma(v_1,v_2,..,v_n) = \frac{1}{N}\sum_{j=1}^N 1-\operatorname{cosine}(\mathbf{e}_{k^*},\mathbf{v}_j)
    \end{equation} 

The caveat of using this is that DDIM's vector $z$ would need to be recomputed and replaced with the VQ-VAE encoding $v$ for each song. However, since the similarity of multiple input songs is only computed during training, this would not slow down our platform.

\begin{table*}[!b]
    \centering
    \caption{Comparison of overall average ratings and counts across versions (v0, v1, and v2) with a single song in the input. The overall row represents the weighted average rating and the total count of songs for each version. Note that count refers to the number of generated songs}
    \begin{tabular}{lcccccc}
        \toprule
        \textbf{Song} & \textbf{Rating (v0)} & \textbf{Rating (v1)} & \textbf{Rating (v2)} & \textbf{Count (v0)} & \textbf{Count (v1)} & \textbf{Count (v2)} \\
        \midrule
        song1  & 3.49 & 3.70 & \textbf{3.90} & 50 & 38 & 21 \\
        song2  & 2.48 & 2.26 & \textbf{2.66} & 52 & 69 & 16 \\
        song3  & 2.99 & 3.41 & \textbf{3.79} & 40 & 51 & 14 \\
        song4  & 2.27 & 2.78 & \textbf{3.05} & 60 & 57 & 11 \\
        song5  & 1.73 & 1.93 & \textbf{2.85} & 24 & 30 & 14 \\
        song6  & 1.66 & 1.63 & \textbf{2.61} & 51 & 37 & 19 \\
        song7  & 3.49 & \textbf{3.82} & 3.78 & 55 & 46 & 23 \\
        song8  & 2.88 & 3.30 & \textbf{3.43} & 62 & 54 & 20 \\
        song9  & 2.78 & \textbf{3.10} & 3.06 & 62 & 59 & 16 \\
        song10 & 3.48 & 3.68 & \textbf{3.80} & 45 & 52 & 20 \\
        \midrule
        \textbf{Overall} & 2.75 & 2.95 & \textbf{3.34} & 501 & 493 & 174 \\
        \bottomrule
    \end{tabular}
    \label{tab:comparison_updated}
\end{table*}

\section{System Architecture} \label{sysarc}
The following describes the architecture and flow of our system.

\begin{itemize}
    \item \textbf{User Interface}: The participant here can log in and start generating the music. These pages also enable the collection of feedback.
    
    \item \textbf{System Logic}: This component controls how the generated songs are displayed to the user. It operates on the dispatch logic mentioned in \ref{sec:algorithm}. Additionally, it oversees song quality control and the collection of training pairs, which are essential for the continuous improvement of the music generation process.

    \item \textbf{Music generator}: This deals with merging and generating the songs along with starting the retraining cycle. 

    \item \textbf{Database}: This stores all the necessary information and music data.
\end{itemize}
The following Figure \ref{fig:architecture} shows the system architecture.

\begin{figure*}[h]
    \centering
    \includegraphics[scale=0.35]{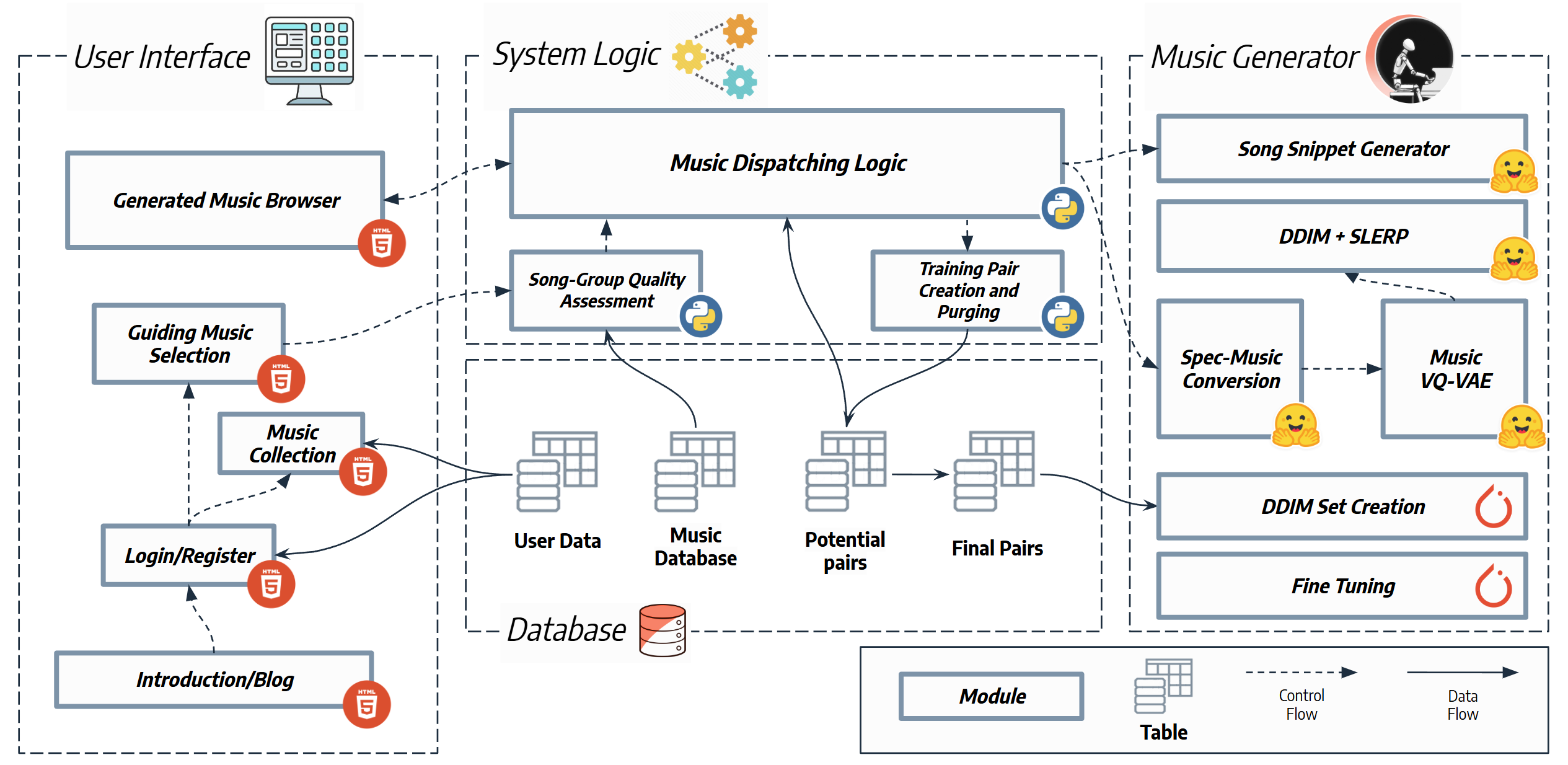}
    \caption{A diagram of the System Architecture.}
    \label{fig:architecture}
\end{figure*}

\section{Pilot Test and Results}
The following section describes the process and results from our pilot study. For the base model, we used the audio diffusion pipeline \cite{audiodiffusion} with teticio/audio-diffusion-256 pre-trained weights. Using a pretrained model allows us to solve the cold start problem in human computation \cite{lika2014facing}. The repository also provides the pipeline for SLERP and noise embedding for our aggregation function. \par
For the first part of our pilot, we worked with 5 selected users. The users initially generated around 50 songs using the base model (v0) and rated them on a scale of 1 to 5, with increments of 0.5. Since our training dataset at this stage is empty, the top 100 songs based on users' normalized ratings were selected to fine-tune the first version (v1).\par
Following this, access to the platform was given to 50 selected users with no restrictions on the number of generated songs. To avoid confirmation bias \cite{suzuki2021characterizing} in our comparison results, the platform would randomly select between version v0 and v1 while generating music. During this, the nearest training sample $z_0$ was also given to the user to collect the baseline rating. \par
To fine-tune the next version v2, we select 100 songs generated from v1 that the users rated better than $rating_{z_0}$. Here, we only select the top 100 to ensure that the improvement is not a factor of a larger training sample. More generations were made with the same group of users, with all three versions to compile the final results. \par
For a single song as input, our results show an incremental increase in the weighted average ratings with each version. On average, v1 improved over v0 by 0.20 points, and v2 improved over v1 by 0.39 points. However, there are examples like songs 2, 7, and 9, where a prior version performed better than its fine-tuned counterpart. The full results can be seen in Table \ref{tab:comparison_updated}. \par
For a combination of two or more songs, the first iterations of the study with 5 users did not yield enough high-rated samples, and therefore, none made it to the dataset used in fine-tuning for v1. Although in our collected data, v1 performs 1.185 points better than v0, the sample size used to calculate this average is too small. The reported reason for this lack is that a combination of multiple songs does not yield distinct-sounding results and is often more distorted. Due to this, we excluded combinations of songs from the last stage of our study.

 
\section{Conclusion}

In conclusion, our proposed human computation approach addresses the challenges faced by traditional music generation models, such as limited input understanding and the absence of consideration for human feedback. Incorporating the ratings as a filtering method and a weight factor in the loss helps the algorithm identify the best training setup and improve the model without the need for additional data. While the pilot test demonstrated encouraging results in terms of our success criteria as described in \ref{successcrit}, the small sample size emphasizes the need for further studies with a larger user base and multiple iterations. This is particularly true for a more complicated problem like the combination of 2 or more song samples to generate music.\par

In the future, implementing the ideas of the purge cycle for continuous refinement, more sophisticated metrics for user feedback, and expanding the user base will contribute to a more comprehensive evaluation of the system. Overall, the integration of human computation in the music generation process presents a promising avenue for achieving personalized and high-quality musical outputs.

\bibliography{hhai}

\begin{thebibliography}{10}

\bibitem{NIPS2014_5ca3e9b1}
Goodfellow I, Pouget-Abadie J, Mirza M, Xu B, Warde-Farley D, Ozair S, et~al.
\newblock Generative Adversarial Nets.
\newblock In: Ghahramani Z, Welling M, Cortes C, Lawrence N, Weinberger KQ, editors. Advances in Neural Information Processing Systems. vol.~27. Curran Associates, Inc.; 2014. Available from: \url{https://proceedings.neurips.cc/paper_files/paper/2014/file/5ca3e9b122f61f8f06494c97b1afccf3-Paper.pdf}.

\bibitem{vaswani2017attention}
Vaswani A, Shazeer N, Parmar N, Uszkoreit J, Jones L, Gomez AN, et~al.
\newblock Attention is all you need.
\newblock Advances in neural information processing systems. 2017;30.

\bibitem{ramesh2021zero}
Ramesh A, Pavlov M, Goh G, Gray S, Voss C, Radford A, et~al.
\newblock Zero-shot text-to-image generation.
\newblock In: International Conference on Machine Learning. PMLR; 2021. p. 8821-31.

\bibitem{yu2022scaling}
Yu J, Xu Y, Koh JY, Luong T, Baid G, Wang Z, et~al.
\newblock Scaling autoregressive models for content-rich text-to-image generation.
\newblock arXiv preprint arXiv:220610789. 2022.

\bibitem{van2017neural}
Van Den~Oord A, Vinyals O, et~al.
\newblock Neural discrete representation learning.
\newblock Advances in neural information processing systems. 2017;30.

\bibitem{rombach2022high}
Rombach R, Blattmann A, Lorenz D, Esser P, Ommer B.
\newblock High-resolution image synthesis with latent diffusion models.
\newblock In: Proceedings of the IEEE/CVF Conference on Computer Vision and Pattern Recognition; 2022. p. 10684-95.

\bibitem{peebles2023scalable}
Peebles W, Xie S.
\newblock Scalable diffusion models with transformers.
\newblock In: Proceedings of the IEEE/CVF International Conference on Computer Vision; 2023. p. 4195-205.

\bibitem{Hill2011MarkovMG}
Hill S.
\newblock Markov Melody Generator; 2011. Available from: \url{https://api.semanticscholar.org/CorpusID:38324438}.

\bibitem{Sherstinsky_2020}
Sherstinsky A.
\newblock Fundamentals of Recurrent Neural Network (RNN) and Long Short-Term Memory (LSTM) network.
\newblock Physica D: Nonlinear Phenomena. 2020 Mar;404:132306.
\newblock Available from: \url{http://dx.doi.org/10.1016/j.physd.2019.132306}.

\bibitem{6795963}
Hochreiter S, Schmidhuber J.
\newblock Long Short-Term Memory.
\newblock Neural Computation. 1997;9(8):1735-80.

\bibitem{roberts2019hierarchical}
Roberts A, Engel J, Raffel C, Hawthorne C, Eck D. A Hierarchical Latent Vector Model for Learning Long-Term Structure in Music; 2019.

\bibitem{dhariwal2020jukebox}
Dhariwal P, Jun H, Payne C, Kim JW, Radford A, Sutskever I.
\newblock Jukebox: A generative model for music.
\newblock arXiv preprint arXiv:200500341. 2020.

\bibitem{engel2019gansynth}
Engel J, Agrawal KK, Chen S, Gulrajani I, Donahue C, Roberts A. GANSynth: Adversarial Neural Audio Synthesis; 2019.

\bibitem{Forsgren_Martiros_2022}
Forsgren S, Martiros H. {Riffusion - Stable diffusion for real-time music generation}; 2022.
\newblock Available from: \url{https://riffusion.com/about}.

\bibitem{huang2023noise2music}
Huang Q, Park DS, Wang T, Denk TI, Ly A, Chen N, et~al.
\newblock Noise2Music: Text-conditioned Music Generation with Diffusion Models.
\newblock arXiv preprint arXiv:230203917. 2023.

\bibitem{schneider2023mousai}
Schneider F, Kamal O, Jin Z, Schölkopf B. Mo\^usai: Text-to-Music Generation with Long-Context Latent Diffusion; 2023.

\bibitem{10.1145/1978942.1979148}
Quinn AJ, Bederson BB.
\newblock Human Computation: A Survey and Taxonomy of a Growing Field.
\newblock In: Proceedings of the SIGCHI Conference on Human Factors in Computing Systems. CHI '11. New York, NY, USA: Association for Computing Machinery; 2011. p. 1403–1412.
\newblock Available from: \url{https://doi.org/10.1145/1978942.1979148}.

\bibitem{dell2024toward}
Dell'Anna D, Murukannaiah PK, Dudzik B, Grossi D, Jonker CM, Oertel C, et~al.
\newblock Toward a quality model for hybrid intelligence teams.
\newblock In: 23rd International Conference on Autonomous Agents and Multiagent Systems, AAMAS 2024. ACM Press Digital Library; 2024. p. 434-43.

\bibitem{mao2024hi}
Mao Y, Rafner J, Wang Y, Sherson J.
\newblock HI-TAM, a hybrid intelligence framework for training and adoption of generative design assistants.
\newblock Frontiers in Computer Science. 2024;6:1460381.

\bibitem{song2020denoising}
Song J, Meng C, Ermon S.
\newblock Denoising diffusion implicit models.
\newblock arXiv preprint arXiv:201002502. 2020.

\bibitem{ho2020denoising}
Ho J, Jain A, Abbeel P.
\newblock Denoising diffusion probabilistic models.
\newblock Advances in neural information processing systems. 2020;33:6840-51.

\bibitem{jafari2014spherical}
Jafari M, Molaei H.
\newblock Spherical linear interpolation and B{\'e}zier curves.
\newblock General Scientific Researches. 2014;2(1):13-7.

\bibitem{kaya2019deep}
Kaya M, Bilge H{\c{S}}.
\newblock Deep metric learning: A survey.
\newblock Symmetry. 2019;11(9):1066.

\bibitem{audiodiffusion}
Smith RD. Audio Diffusion. GitHub; 2024.
\newblock \url{https://github.com/teticio/audio-diffusion}.

\bibitem{lika2014facing}
Lika B, Kolomvatsos K, Hadjiefthymiades S.
\newblock Facing the cold start problem in recommender systems.
\newblock Expert systems with applications. 2014;41(4):2065-73.

\bibitem{suzuki2021characterizing}
Suzuki M, Yamamoto Y.
\newblock Characterizing the influence of confirmation bias on web search behavior.
\newblock Frontiers in psychology. 2021;12:771948.

\end{thebibliography}
\end{document}